\documentclass[runningheads,citeauthoryear]{apinv}
\usepackage{epsfig,cite,graphics}

\usepackage[T2A]{fontenc}
\usepackage[cp1251]{inputenc}
\usepackage[bulgarian,english]{babel}

\begin{document}

\title{The FUor Candidate V582 Aurigae: First Photometric and Spectroscopic Observations}
\titlerunning{The FUor Candidate V582 Aur}
\author{Evgeni H. Semkov\inst{1}, Stoyanka P. Peneva\inst{1}, Michel Dennefeld\inst{2}}
\authorrunning{Semkov et al.}
\tocauthor{Evgeni H. Semkov, Stoyanka P. Peneva, Michel Dennefeld}
\institute{Institute of Astronomy, Bulgarian Academy of Sciences, Sofia, Bulgaria
        \and Institut d'Astrophysique de Paris, CNRS, and Universit\'{e} P. et M. Curie, Paris, France 
        \newline
        \email{esemkov@astro.bas.bg, speneva@astro.bas.bg}
}
\papertype{poster}
\maketitle

\begin{abstract}
One of the most attractive events in the pre-main sequence evolution is the FU Orionis (FUor) outburst. 
Because only a small number of FUor stars have been detected to date, photometric and spectral studies of every new object are of great interest. Recently, a new FUor candidate was discovered by Anton Khruslov - V582 Aur. 
To confirm the FUors nature of this object we started regular photometric observations with the telescopes of the National Astronomical Observatory Rozhen (Bulgaria). 
A high-resolution spectrum of V582 Aur was obtained with the 1.93 m telescope in Haute-Provence Observatory (France). 
\end{abstract}


\Bg
\bgtitle{Кандидатът за FUor обект V582 Aurigae: първи фотометрични и спектрални наблюдения}  
{Стоянка П. Пенева, Евгени Х. Семков, Мишел Денефелд}  
{Едно от най-атрактивните явления по време на еволюцията на звездите преди Главната последователност са избухванията от типа FU Orionis (FUor). Тъй като до сега са известни само няколко обекта от този клас, всеки нов подобен обект предизвиква сериозен интерс за фотометрични и спектрални изследвания. Неотдавна, нов предполагаем обект от типа FUor беше открит от Антон Хруслов - V582 Aur. За да потвърдим принадлежността на V582 Aur към FUor променливите ние започнахме редовни фотометрични наблюдения с телескопите на НАО - Рожен (България). С 1.93-метровият телескоп на обсерваторията в Горен Прованс (Франция) беше получен спектър с висока дисперсия на V582 Aur.}
\Eng

\section{Introduction}
FUors are low-mass Pre-main sequence (PMS) stars defined as a class by Herbig (1977) after the discovering of V1057 Cyg and V1515 Cyg. 
The main characteristics of FUors are an increase in optical brightness of about 4-5 mag, a F-G supergiant spectrum with broad blue-shifted Balmer lines, strong infrared excess, connection with reflection nebulae, and location in star-forming regions (Reipurth 1990, Bell et al. 1995, Clarke et al. 2005). 
Typical spectroscopic properties of FUors include a gradual change in the spectrum from earlier to later spectral type from the blue to the infrared, a strong Li I ($\lambda$ 6707) line, P Cygni profiles of H$\alpha$ and Na I ($\lambda$ 5890/5896) lines, and the presence of CO bands in the near infrared spectra (Herbig 1977, Bastian \& Mundt 1985). 
The prototypes of FUors seem to be low-mass PMS objects (T Tauri stars) with massive circumstellar disks. 
According to a commonly accepted view, the FUor outburst is produced by a sizable increase in accretion from a circumstellar disk on the stellar surface. 
The cause of this increase in accretion from $\sim$10$^{-7}$$M_{\sun}$$/$yr to $\sim$10$^{-4}$$M_{\sun}$$/$yr appears to be thermal or gravitational instability in the circumstellar disk (Hartmann \& Kenyon 1996).

Recently, a new FUor candidate was discovered by Anton Khruslov. 
The variable was catalogued as USNO A2.0 1200-03303169. 
Samus (2009) compares the brightness of star on the photographic plates in Moscow collection (1965-1992) and on the images from Digitized Sky Survey (1954-1993) suspects that the brightening started between 1982 and 1986. 
The star has been added to the General Catalogue of Variable Stars with the designation V582 Aur (Samus 2009). 
Munari et al. (2009) obtained a low-resolution CCD spectrum of V582 Aur on Aug. 6, 2009 with the 1.22-m telescope of the Asiago Astrophysical Observatory (Italy). 
The authors registered the presence of absorption lines of the Balmer series, Na I D and Ba II ($\lambda$ 6496) and absence of the Li I ($\lambda$ 6707) line in the spectrum. 
The photometric observations of V582 Aur reported by Munari et al. (2009) show the star near the maximal brightness on Aug. 5 and 6, 2009.

\section{Observations}

Our photometric observations were obtained with the 2-m RCC and 50/70-cm Schmidt telescopes of the National Astronomical Observatory Rozhen (Bulgaria). 
We used the Vers Array 1300B CCD camera with the 2-m RCC telescope and the FLY PL 16803 CCD camera with the 50/70-cm Schmidt telescope. 
All frames were taken through a standard Johnson-Cousins set of filters.
Aperture photometry was performed using IDL DAOPHOT routines.
The procedure used calculates the centroid of stellar object and the mean value of the background around it.
The digitized plates from the Palomar Schmidt telescope, available via the website of the Space Telescope Science Institute, are used, too.

A spectrum of V582 Aur was obtained on Jan 15, 2010 with the SOPHIE echelle spectrograph installed on the 1.93 m telescope at the Haute-Provance Observatory (France). 
We used the high-efficiency mode with spectral resolution of R $\sim$ 45 000. 
The wavelength range is 3872-6943 \AA, exposure time is 50 min, attaining signal-to-noise ratio S/N=20 around H$\alpha$. 
The instrument is entirely computer controlled and a standard reduction of data is done automatically. 
This reduction includes bias subtraction, optimal order extraction, cosmic ray removal, flat-fielding, wavelength calibration, cross-correlation with a suitable numerical mask and merging of the spectral orders. 
The spectrum has been normalized by fitting a spline function to continuum points with IRAF task CONTIN. 
The measuring of spectral lines width was done with IRAF task SPLOT by fitting Gaussian functions to spectral lines. 

\section{Results}

Our photometric and spectroscopic study confirms the affiliation of V582 Aur to the group of FUor objects. 
In Fig. 1 we plot $BVRI$ light curves of V582 Aur from all available photometric observations. 
The magnitudes calibration of our data was carried out with respect to nine stars from USNO-B1.0 Catalog (Monet et al. 2003). For POSS-I and QUICK-V observations these stars are: 1248-0097550, 1248-0097575, 1248-0097497, 1248-0097478 and 1248-0097467, and for POSS-II and our CCD observations: 1248-0097490, 1248-0097541, 1248-0097599 and 1248-0097579. 
Due to the different standard stars used for photometry, differences between our results and published in Samus (2009) and Munari et al. (2009) can be expected. 
Nevertheless, an outburst in optical light and a rise in brightness in the period 1982-2010 are well documented. 
During this period, the increase in brightness passed slowly, and the V-band magnitude increased by 4$^m$.

\begin{figure}[!htb]
  \begin{center}
    \centering{\epsfig{file=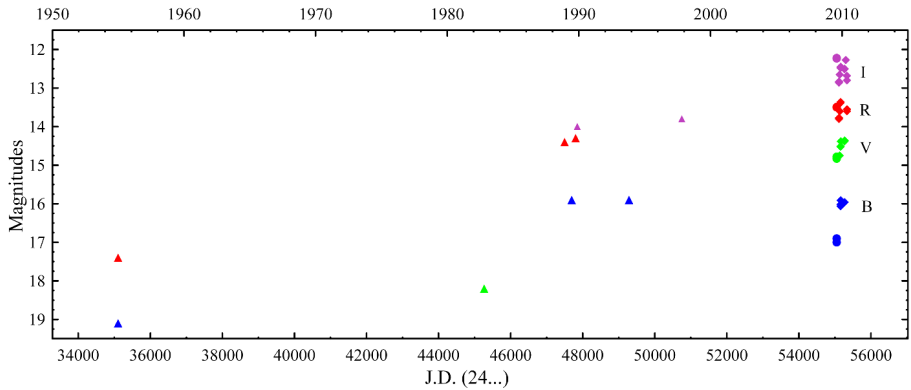}}
    \caption[]{$BVRI$ light curves of V582 Aur in the period 1954 – 2010. The diamonds denote data from our CCD observations, the circles: CCD data from Munari et al. (2009), the triangles: magnitudes, obtained from POSS-I, POSS-II and Quick-V plates.}
    \label{countryshape}
  \end{center}
\end{figure}

In Table 1 the identification of absorption lines in the spectrum of V582 Aur together with equivalent widths (EW) are listed. 
The H$\alpha$ line and NaI doublet show the typical for FUor stars P Cygni profiles (Fig. 2). 
The broad blueshifted H$\alpha$ absorption seems to be saturated, extending to about -800 km/s that is produced by powerful, rapidly expanding wind.
The absence of Li I ($\lambda$ 6707) in our spectrum may be explained by the very strong variability of this line in the FUor spectra (Herbig 2009). 

 \begin{table}[htb]
  \begin{center}
  \caption{Equivalent widths (EW) for the absorption lines in spectrum of V582 Aur}
  \begin{tabular}{llllll}
  \hline
  \noalign{\smallskip}   
Line   & & $\lambda$ [\AA]  & && 	EW [\AA]\\
 \noalign{\smallskip}
  \hline
  \noalign{\smallskip}
H$\beta$ & & 4861.33 & &&	5.7\\
Na I     & & 5889.95 & &&	2.1\\
Na I     & & 5895.92 & &&	1.9\\
H$\alpha$& & 6562.82 & &&  8.4\\
Ba II    & & 6495.68 & &&	1.6\\
\noalign{\smallskip}
\hline
 \end{tabular}
  \label{table1}
  \end{center}
 \end{table}

The light curve of V582 Aur in the period of increase in brightness is similar to that observed for the classical FUor star V1515 Cyg, but the time of rise seems to be longer and it continue up to now.
It is currently impossible to determine the time of the beginning of the optical outburst due to a lack of photometric data for the period 1982-1986 (Samus 2009).
One possibility for the correct determination of the outburst parameters and construction of the historical light curve of FUors is a search in the photographic plate archives (Peneva et al. 2010). 
Our preliminary search in the Wide Field Plates Data Base (Tsvetkov et al. 1997) shows the possibility to find archival photographic observations of V582 Aur in the plate collections of the following telescopes:
the 67/92 cm Schmidt telescope at Asiago Observatory (Italy),  
the 105/150 cm Schmidt telescope at Kiso Observatory (Japan),
the 60/90 cm Schmidt telescope at Konkoly Observatory (Hungary), 
and the 134/200 cm Schmidt telescope at Tautenburg Observatory (Germany).
We plan to calibrate a sequence of photometric standards in the field of V582 Aur and use it to measure all available photometric observations.

\begin{figure}[!htb]
  \begin{center}
    \centering{\epsfig{file=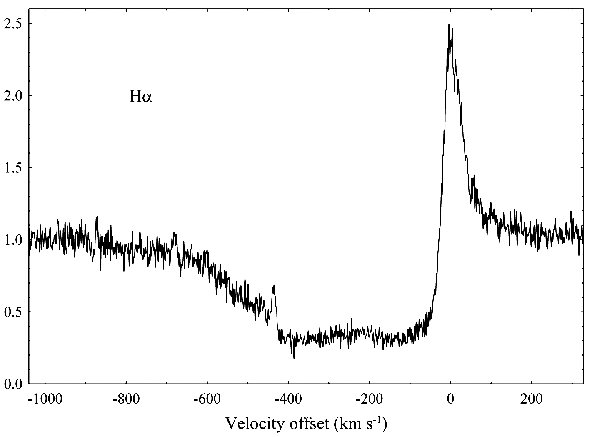, width=0.45\textwidth}
    {\epsfig{file=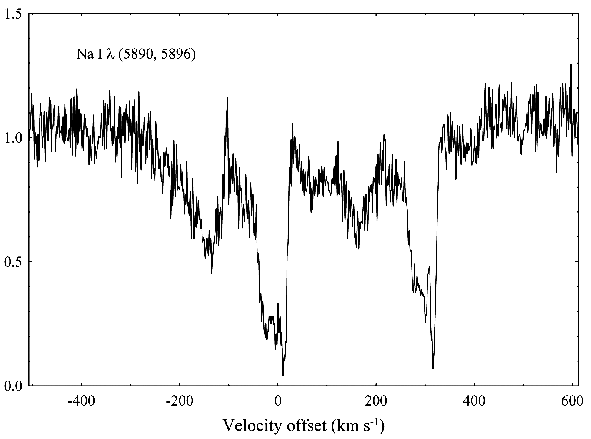, width=0.45\textwidth}}}
    \caption[]{The line profiles of H$\alpha$, and Na I doublet from a high-resolution spectrum of V582 Aur, taken with the SOPHIE echelle spectrograph, installed at the 1.93 m telescope in Haute-Provence Observatory}
    \label{countryshape}
  \end{center}
\end{figure}

{\it Acknowledgments:} This work was partly supported by grants DO 02-85, DO 02-273 and DO 02-362 of the National Science Fund of the Ministry of Education, Youth and Science, Bulgaria. 
The Digitized Sky Survey was produced at the Space Telescope Science Institute under U.S. Government grant NAG W-2166. The images of these surveys are based on photographic data obtained using the Oschin Schmidt Telescope on Palomar Mountain and the UK Schmidt Telescope. 
The plates were processed into the present compressed digital form with the permission of these institutions. This research has made use of the NASA’s Astrophysics Data System.            


\end{document}